\documentclass[aps,pra,reprint,floatfix,superscriptaddress,fleqn,nofootinbib]{revtex4-2}
\usepackage{mathtools,amssymb,amsthm,bm,bbm,xcolor,mathdots,stmaryrd}
\usepackage[colorlinks=true, linkcolor=blue, citecolor=magenta, urlcolor=blue]{hyperref}
\pdfoutput=1

\usepackage[T1]{fontenc}
\usepackage{lmodern}
\usepackage[utf8]{inputenc}

\usepackage{microtype}
\usepackage{orcidlink}

\usepackage[caption=false]{subfig}
\usepackage{graphicx}

\theoremstyle{remark}

\newcommand{\tr}{\operatorname{tr}}

\newcommand{\bra}[1]{\langle #1|}
\newcommand{\ket}[1]{|#1\rangle}

\newcommand{\ketbra}[2]{|#1 \rangle\langle #2|}

\newcommand{\tauMT}{\tau_\textsc{mt}}
\newcommand{\tauclMT}{\bar\tau_\textsc{mt}}
\newcommand{\tauML}{\tau_\textsc{ml}}
\newcommand{\tauBD}{\tau_\textsc{bd}}
\newcommand{\tauclBD}{\bar\tau_\textsc{bd}}

\newcommand{\llangle}{\langle\hspace{-2pt}\langle}
\newcommand{\rrangle}{\rangle\hspace{-2pt}\rangle}

\newcommand{\emax}{\epsilon_\text{max}}
\newcommand{\emin}{\epsilon_\mathrm{min}}
\newcommand{\emint}{\epsilon_{\text{min};t}}
\newcommand{\etmint}{\epsilon_{\text{min},\theta;t}}
\newcommand{\emaxt}{\epsilon_{\text{max};t}}

\begin{document}
\title{Closed systems refuting quantum-speed-limit hypotheses}
\author{Niklas H{\"o}rnedal\,\orcidlink{0000-0002-2005-8694}\,}
\email{niklas.hornedal@uni.lu}
\affiliation{Department of Physics and Materials Science, University of Luxembourg, L-1511 Luxembourg, Luxembourg}
\author{Ole S{\"o}nnerborn\,\orcidlink{0000-0002-1726-4892}\,}
\email{ole.sonnerborn@kau.se}
\affiliation{Department of Mathematics and Computer Science, Karlstad University, SE-651\,88 Karlstad, Sweden}
\affiliation{Department of Physics, Stockholm University, SE-106\,91 Stockholm, Sweden}

\begin{abstract}
Many quantum speed limits for isolated systems can be generalized to also apply to closed systems. This is, for example, the case with the well-known Mandelstam-Tamm quantum speed limit. Margolus and Levitin derived an equally well-known and ostensibly related quantum speed limit, and it seems to be widely believed that the Margolus–Levitin quantum speed limit can be similarly generalized to closed systems. However, a recent geometrical examination of this limit reveals that it differs significantly from most known quantum speed limits. In this paper, we show that, contrary to the common belief, the Margolus-Levitin quantum speed limit does not extend to closed systems in an obvious way. More precisely, we show that for every hypothetical bound of Margolus-Levitin type, there are closed systems that evolve with a conserved normalized expected energy between states with any given fidelity in a time shorter than the bound. We also show that for isolated systems, the Mandelstam-Tamm quantum speed limit and a slightly weakened version of this limit that we call the Bhatia-Davies quantum speed limit always saturate simultaneously. Both of these evolution time estimates extend straightforwardly to closed systems. We demonstrate that there are closed systems that saturate the Mandelstam-Tamm but not the Bhatia-Davies quantum speed limit.
\end{abstract}
\date{\today}
\maketitle

\section{Introduction}
Many quantum speed limits (QSLs) for isolated systems can be generalized to also apply to closed systems \cite{PiCiCeAdSo-Pi2016, Fr2016, DeCa2017}. By an isolated system we mean one whose Hamiltonian does not change over time, while a closed system may have a time-varying Hamiltonian. The famous Mandelstam-Tamm QSL is of this kind \cite{MaTa1945, AnAh1990}. The Mandelstam-Tamm QSL states that 
it takes at least the time 
\begin{equation}\label{MT}
    \tauMT
    = \frac{\pi}{2\Delta H}
\end{equation}
for an isolated system to evolve between two fully distinguishable states.\footnote{``State'' will always refer to a pure quantum state, that is, a state that can be represented by a density operator of rank $1$. Two states $\rho_1$ and $\rho_2$ are fully distinguishable if their fidelity vanishes.}\textsuperscript{,}\footnote{All quantities are expressed in units such that $\hbar=1$.} Here, $\Delta H$ is the energy uncertainty. More generally, it takes at least the time 
\begin{equation}\label{isolatedMT}
	\tauMT(\delta)
	= \frac{\arccos\sqrt{\delta}}{\Delta H}
\end{equation}
for an isolated system to evolve between two states with fidelity $\delta$.\footnote{The fidelity between two states $\rho_1$ and $\rho_2$ is $\tr(\rho_1\rho_2)$.
} The originators of the estimate $\tauMT(\delta)$ are also Mandelstam and Tamm, but it was rediscovered and formulated in a more concise way in Ref.\ \cite{Fl1973}.

The Mandelstam-Tamm QSL can be extended to closed systems by replacing the denominator in \eqref{isolatedMT} with the corresponding time average. Thus, the evolution time of a closed system evolving between two states with fidelity $\delta$ is bounded from below by 
\begin{equation}\label{closedMT}
	\tauclMT(\delta)
	= \frac{\arccos\sqrt{\delta}}{\llangle\Delta H_t\rrangle},
\end{equation}
with $\llangle\Delta H_t\rrangle$ being the time average of the energy uncertainty. Since the Fubini-Study distance between two states with fidelity $\delta$ is $\arccos\sqrt{\delta}$, and the Fubini-Study speed with which a state evolves is $\Delta H_t$ \cite{AnAh1990}, the Mandelstam-Tamm QSL is saturated if and only if the state follows a Fubini-Study geodesic in the projective Hilbert space. Mandelstam and Tamm's QSL has also been extended to systems in mixed states \cite{Uh1992, AnHe2014, HoAlSo2022}.

Margolus and Levitin \cite{MaLe1998} derived a seemingly similar evolution time estimate. The Margolus-Levitin QSL states that the time it takes for an isolated system to evolve between two fully distinguishable states is greater than or equal to
\begin{equation}\label{ML}
    \tauML=\frac{\pi}{2\langle H-\emin\rangle},
\end{equation}
where $\langle H-\emin\rangle$ is the expected energy $\langle H\rangle$ shifted by the smallest occupied energy $\emin$.\footnote{An energy $\epsilon$ is occupied by $\rho$ if $\bra{\epsilon}\rho\ket{\epsilon}>0$ for an energy eigenstate $\ket{\epsilon}$ with eigenvalue $\epsilon$.} The quantity $\langle H-\emin\rangle$ is hereafter referred to as the normalized expected energy. A more general result states that the time it takes for an isolated system to evolve between two states with fidelity $\delta$ is lower bounded by
\begin{equation}\label{extML}
    \tauML(\delta)
    =\frac{\alpha(\delta)}{\langle H-\emin\rangle},
\end{equation}
where 
\begin{equation}\label{alpha}
    \alpha(\delta)
    =\min_{z^2\leq \delta}\bigg\{\frac{1+z}{2}\arccos\bigg(\frac{2\delta-1-z^2}{1-z^2}\bigg)\bigg\}.
\end{equation}
As $\tauMT(\delta)$, the bound $\tauML(\delta)$ is tight, and $\tauML(0)=\tauML$. The bound $\tauML(\delta)$ was established numerically in Ref.\ \cite{GiLlMa2003} and derived analytically in Ref.\ \cite{HoSo2023}. Reference \cite{HoSo2023} also contains a geometric interpretation of $\tauML(\delta)$ and a complete description of the systems that reach the bound.

A natural guess is that the Margolus-Levitin QSL is also valid for closed systems provided the time average of the normalized expected energy is placed in the denominator. More generally, one might expect that there is a QSL for closed systems of the form $\mathcal{L}(\delta)/\llangle H_t-\emint\rrangle$ where $\mathcal{L}$ is some non-negative function that depends only on the fidelity $\delta$ between the initial and final states and $\llangle H_t-\emint\rrangle$ is the time average of the instantaneous normalized expected energy.\footnote{To be interesting, $\mathcal{L}$ in such a hypothetical QSL should satisfy $\mathcal{L}(1)=0$ and $\mathcal{L}(\delta)>0$ for $0\leq\delta<1$.} It has been argued that one can take $\mathcal{L}(\delta)=\arccos\sqrt{\delta}$ \cite{DeLu2013a,DeLu2013b,SuZh2019}. However, for $0<\delta<1$, such a hypothetical QSL is violated by an isolated system that evolves between two states with fidelity $\delta$ in time $\tauML(\delta)$. This is because $\alpha(\delta)$ is strictly smaller than $\arccos\sqrt{\delta}$ for $0<\delta<1$ \cite{GiLlMa2003,HoSo2023}. For $\delta=0$, such a hypothetical QSL can be violated by a closed system in the family of systems constructed in the next section.\footnote{Reference \cite{DeLu2013a} has been criticized before \cite{OkOh2018}. However, the modified Margolus-Levitin estimate in Ref.\ \cite{OkOh2018} is also incorrect because it is violated by any isolated system that evolves between two states with fidelity $0<\delta <1$ in time $\tauML(\delta)$ \cite{HoSo2023}.} It has also been argued that one can take $\mathcal{L}(\delta)=\sin^2(\arccos\sqrt{\delta})/2=(1-\delta)/2$ \cite{DeLu2013b}. Nor is that possible as we show below. 

The main result of this paper is as follows: For each state $\rho$, each fidelity $\delta$, and each positive $\mathcal{L}(\delta)$, there exists a time-dependent Hamiltonian $H_t$ that evolves $\rho$ to a state with fidelity $\delta$ relative to $\rho$ in a time less than $\mathcal{L}(\delta)/\llangle H_t-\emint\rrangle$. The Hamiltonian can be chosen such that the normalized expected energy remains fixed at an arbitrary predetermined value.

Lui \emph{et al.}\ \cite{LiMiFuWa2021} used the Bhatia–Davies inequality to transform the Mandelstam-Tamm QSL into an upper bound for an operationally defined QSL \cite{ShLiZhYuLi2020}. This upper bound is a new QSL that we call the Bhatia-Davies QSL, although we should rightly attribute it to the authors of Ref.\ \cite{LiMiFuWa2021}. The Bhatia-Davies QSL states that the time it takes for an isolated system to evolve between two states with fidelity $\delta$ is bounded from below by
\begin{equation}\label{isolatedBD}
    \tauBD(\delta) = \frac{\arccos\sqrt{\delta}}{\sqrt{\langle\emax -  H\rangle\langle H - \emin\rangle}},
\end{equation}
where $\emax$ is the largest and $\emin$ is the smallest occupied energy. The Bhatia-Davies QSL also extends straightforwardly to closed systems:
\begin{equation}\label{closedBD}
	\tauclBD(\delta)
    = \frac{\arccos\sqrt{\delta}}{\llangle\sqrt{\langle\emaxt -  H_t\rangle\langle H_t- \emint\rangle}\,\rrangle}.
\end{equation}

The Bhatia-Davies QSL is weaker than that of Mandelstam and Tamm in the sense that $\tauclMT(\delta)\geq \tauclBD(\delta)$ with a strict inequality in general for both isolated and closed systems. We show that the Mandelstam-Tamm and the Bhatia-Davies QSLs are always saturated simultaneously for isolated systems, but that this need not be the case for closed systems: We give examples of closed systems that saturate the Mandelstam-Tamm QSL but not the Bhatia-Davies QSL.

\section{Time-dependent systems that disprove common belief}
One obtains a relatively simple type of time-dependent Hamiltonian if one conjugates a time-independent Hermitian operator $H$ with a one-parameter group of unitaries generated by a Hermitian operator $A$:
\begin{equation}
    H_t=e^{-iAt}H e^{iAt}.
\end{equation}
Such a group action will preserve the eigenvalues but rotate the eigenvectors of $H$. If a state $\rho$ evolves under the influence of $H_t$, that is, if 
\begin{equation}
    \dot\rho_t=-i[H_t,\rho_t],\qquad \rho_0=\rho,
\end{equation}
the state in the rotating frame picture,
\begin{equation}
    \rho^{\textsc{rf}}_t=e^{iAt} \rho_t e^{-iAt},
\end{equation}
evolves as if $H-A$ governs the dynamics:
\begin{equation}\label{rotating}
    \dot\rho^{\textsc{rf}}_t=-i[H-A,\rho^{\textsc{rf}}_t],\qquad \rho^{\textsc{rf}}_0=\rho.
\end{equation}
As a consequence, in the Schrödinger picture, 
\begin{equation}\label{evolution}
    \rho_t=e^{-iAt} e^{-i(H-A)t} \rho e^{i(H-A)t} e^{iAt}.
\end{equation}

The behavior of $\rho_t$ can be quite complex even though $H_t$ has a relatively simple time dependence. However, Eq.\ \eqref{evolution} tells us that if the initial state $\rho$ commutes with $H-A$, the evolving state will behave as if the time-independent ``effective Hamiltonian'' $A$ generates it:
\begin{equation}
    \rho_t=e^{-iAt} \rho e^{iAt}.
\end{equation}
This observation will be of central importance below. 

The eigenvectors of $H_t$ will also evolve with $A$ as an effective Hamiltonian: If $\ket{j}$ is an eigenvector of $H$ with the eigenvalue $\epsilon_j$, then $\ket{j;t}=e^{-iAt}\ket{j}$ is an eigenvector of $H_t$ with the eigenvalue $\epsilon_j$. As a result, the occupations of the energy levels are constant over time:
\begin{equation}
\label{occupied}
    \bra{j;t}\rho_t\ket{j;t}=\bra{j}\rho\ket{j}.
\end{equation}
This means that the expected energy $\langle H_t\rangle$, the energy uncertainty $\Delta H_t$, the normalized expected energy $\langle H_t-\emint\rangle$, and the ``dual'' normalized expected energy $\langle \emaxt-H_t\rangle$ are conserved quantities; see Refs.\ \cite{NeAlSa2022,HoSo2023} for a QSL involving the dual of the normalized expected energy. 

Another important fact is that if the initial state $\rho$ satisfies $A\rho+\rho A=A$, the evolving state $\rho_t$ is a Fubini-Study geodesic; see Appendix A in Ref.\ \cite{HoAlSo2022}. If such is the case, the Mandelstam-Tamm QSL is saturated, and the system evolves between two states with fidelity $\delta$ in time $\tauclMT(\delta)$. Interestingly, given any initial state $\rho$ and any Hermitian operator $H$, there is an elegant way to construct an $A$ such that $[H-A,\rho]=0$ and $A\rho+\rho A=A$: Write $\rho=\ketbra{u}{u}$ and define
\begin{equation}\label{elegant}
    A=(H-\bra{u}H\ket{u})\ketbra{u}{u}+\ketbra{u}{u}(H-\bra{u}H\ket{u}).
\end{equation}
In the following, we show how to disprove two hypotheses about QSLs with appropriate choices of $\rho$ and $H$, and $A$ defined as in \eqref{elegant}.

\subsection{The nonexistence of a time-dependent Margolus-Levitin quantum speed limit}\label{sec:No timeMLQSL}
The Mandelstam-Tamm and Margolus-Levitin QSLs tell us that an isolated system with large energy uncertainty that evolves along a Fubini-Study geodesic has a correspondingly large normalized expected energy \cite{HoSo2023}. Interestingly, this need not hold for systems with time-dependent Hamiltonians. Below we give an example of a family of closed systems evolving along geodesics, whose normalized expected energies are conserved with an arbitrary common predetermined value and whose energy uncertainties are conserved and form an unbounded set.

Fix an $E>0$ and consider a quantum system in a state $\rho=\ketbra{u}{u}$. Let $H_\theta$ be a Hermitian operator, to be specified, and define $A_\theta$ as in Eq.\ \eqref{elegant}:
\begin{equation}
    A_\theta=(H_\theta-\bra{u}H_\theta\ket{u})\ketbra{u}{u}+\ketbra{u}{u}(H_\theta-\bra{u}H_\theta\ket{u}).
\end{equation}
Furthermore, let $H_{\theta;t}=e^{-iA_\theta t}H_\theta e^{iA_\theta t}$ and let $\rho_{\theta;t}$ be the state at time $t$ generated from $\rho$ by $H_{\theta;t}$. Then $\rho_{\theta;t}$ is a Fubini-Study geodesic.

\begin{figure}[t]
	\centering
	\includegraphics[width=0.8\linewidth]{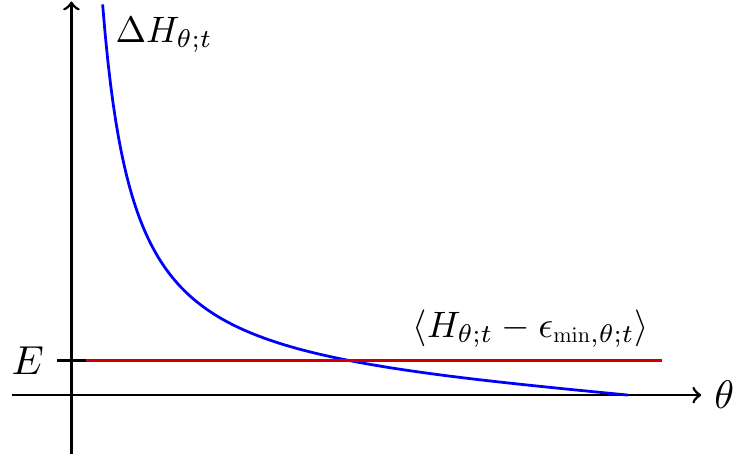}
	\caption{Graphs illustrating the dependence of the energy uncertainty (blue) and the normalized expected energy (red) on the angle $\theta$. The requirement that the normalized expected energy be constant forces the energy uncertainty to grow toward infinity with decreasing angle.}
	\label{fig0}
\end{figure}
To specify $H_\theta$ let $\ket{v}$ be a unit vector perpendicular to $\ket{u}$ and define the Pauli operators $X$ and $Z$ as
\begin{align}
    X &= \ketbra{u}{u}-\ketbra{v}{v}, \label{X}\\
    Z &= \ketbra{u}{v}+\ketbra{v}{u}. \label{Z}
\end{align}
Let $\mu(\theta)=E/(1-\cos\theta)$ for $0 < \theta <\pi$, and define 
\begin{equation}
    H_\theta=\mu(\theta)(\sin\theta Z - \cos\theta X).
\end{equation}
The largest and the smallest eigenvalues of $H_\theta$, and thus of $H_{\theta;t}$, are $\mu(\theta)$ and $-\mu(\theta)$, respectively, both of which are occupied by $\rho_{\theta;t}$. Furthermore, the normalized expected energy and the energy uncertainty are 
\begin{align}
    &\langle H_{\theta;t} - \etmint\rangle = \mu(\theta)(1-\cos\theta)=E, \label{energi} \\
    &\Delta H_{\theta;t} = \mu(\theta)\sin\theta=E\cot(\theta/2). \label{energiosakerhet}
\end{align}
The blue graph in Fig.\ \ref{fig0} illustrates how the energy uncertainty depends on $\theta$, and the red graph accentuates that the normalized expected energy does not depend on $\theta$. Note that both quantities are independent of $t$.
\begin{figure}[t]
	\centering
	\includegraphics[width=0.8\linewidth]{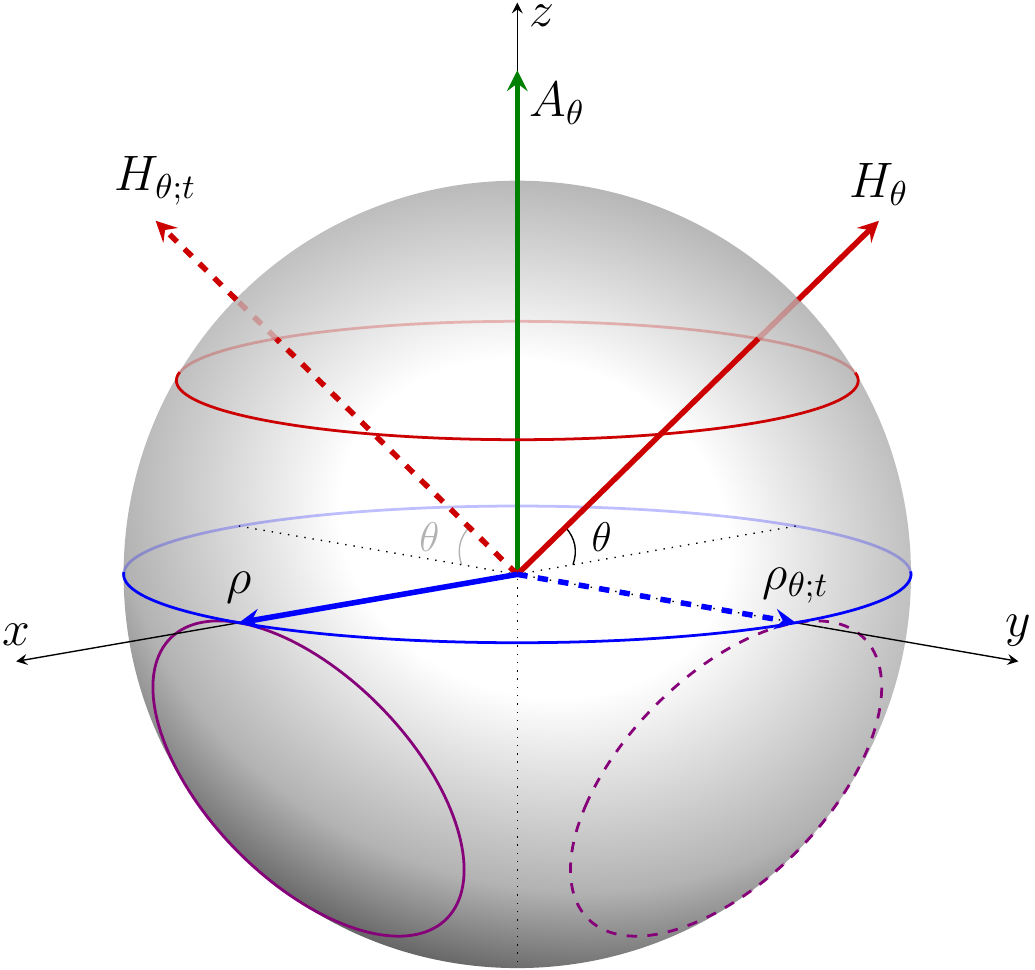}
	\caption{Rabi and Bloch vector representations of $H_\theta$, $A_\theta$, and $\rho$. The vector representing $\rho$ points along the positive $x$ axis, and the vector representing $H_\theta$ makes the angle $\theta$ with the negative $x$ axis. The purple circle represents the expected energy level to which $\rho$ belongs. As time passes, the state and the Hamiltonian rotate around the $z$ axis with the same angular velocity. The dashed vectors represent $H_{\theta;t}$ and $\rho_{\theta;t}$ at a time $t>0$. The expected energy level rotates with the state.}
	\label{fig1}
\end{figure}

Next, fix a fidelity $0\leq\delta<1$ (for $\delta=1$ there is nothing to prove), let $\mathcal{L}(\delta)$ be any positive number representing the numerator in a hypothetical extension of the Margolus-Levitin QSL, and choose $\theta$ such that
\begin{equation}\label{storre}
    \cot(\theta/2)>\frac{\arccos{\sqrt{\delta}}}{\mathcal{L}(\delta)}.
\end{equation}
Furthermore, let $\tau(\delta)$ be the first time $t$ the fidelity between $\rho_{\theta;t}$ and $\rho$ is $\delta$.\footnote{Since $\rho_{\theta,t}$ is a geodesic, the fidelity between $\rho_{\theta,t}$ and $\rho$ will sooner or later be $\delta$.} Since the state follows a Fubini-Study geodesic, the Mandelstam-Tamm QSL is saturated: $\tau(\delta)=\tauclMT(\delta)$. According to Eqs.\ \eqref{energi}--\eqref{storre}, 
\begin{equation}\label{sista}
    \tau(\delta)
    =\frac{\arccos\sqrt{\delta}}{E\cot(\theta/2)}
    <\frac{\mathcal{L}(\delta)}{E}
    =\frac{\mathcal{L}(\delta)}{
    \llangle H_{\theta;t}-\etmint\rrangle}.
\end{equation}
We conclude that for each state $\rho$ and each fidelity $\delta$, there is a time-dependent Hamiltonian that evolves $\rho$ to a state with fidelity $\delta$ relative to $\rho$ in a time less than a hypothetical QSL of the form given by the rightmost expression in \eqref{sista}.
Thus, the Margolus-Levitin QSL does not straightforwardly extend to closed systems.

\vspace{2pt}
In Fig.\ \ref{fig1} we have represented $H_\theta$, $A_\theta$, and $\rho$ as Rabi and Bloch vectors relative to $X$, $Y$, and $Z$, with $Y=i(\ketbra{u}{v}-\ketbra{v}{u})$. The angle between $H_\theta$ and the negative $x$ axis is $\theta$. As time passes, the state and the Hamiltonian rotate around the $z$ axis with the same angular speed. Note that $\rho_{\theta;t}$ moves along the equator in the Bloch sphere and thus is a Fubini-Study geodesic. The dotted vectors represent the state and the Hamiltonian at a time $t>0$.

The purple circle formed by intersecting the Bloch sphere with a plane perpendicular to the extension of the vector representing $H_\theta$ represents the expected energy level to which $\rho$ belongs. This circle rotates together with $H_{\theta;t}$ and always lies in a plane perpendicular to the vector representing $H_{\theta;t}$. The key observation is that this circle corresponds to the normalized expected energy $E$ irrespective of the value of angle $\theta$, and $\rho_{\theta;t}$ will evolve together with that circle.

\begin{figure}[t]
	\centering
	\includegraphics[width=0.9\linewidth]{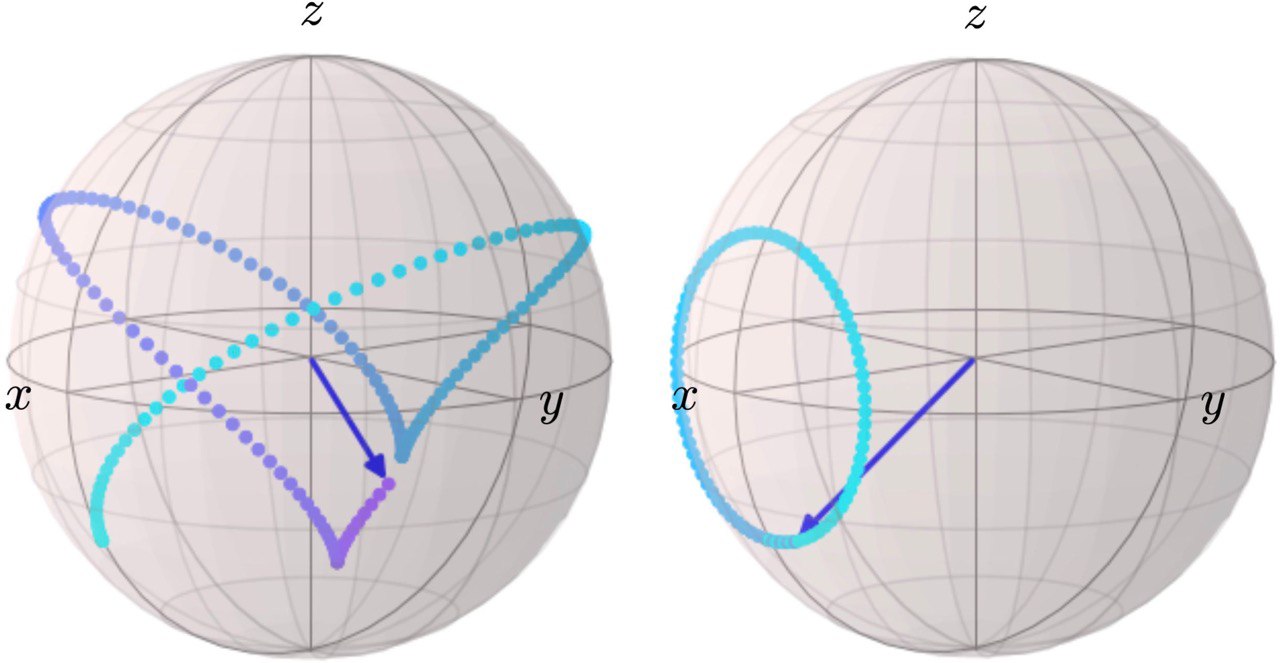}
	\caption{An evolution curve starting from a state not on the equator of the Bloch sphere. Here, $\theta=30^\circ$ and $E=1$. The left figure shows the evolution curve in the Schrödinger picture, and the right figure shows the same curve in the rotating frame picture. The warmer colors indicate more recent times, and the blue arrow represents the state at the final time.}
	\label{fig2}
\end{figure}

Most initial states will not evolve in such a well-behaved manner as those located on the $x$ axis of the Bloch sphere. In Fig.\ \ref{fig2} we have drawn the evolution curve of a state not on the $x$ axis.
In the rotating frame picture \eqref{rotating}, the evolution curve forms a circle around the $x$ axis. 
This is since $H_\theta-A_\theta\propto X$.

\subsection{The Bhatia-Davies quantum speed limit}
In the previous section, we exploited that for some members of a family of closed systems with a common conserved normalized expected energy, the energy uncertainty is large enough for the evolution time to violate any hypothetical straightforward extension of the Margolus-Levitin QSL. For the systems in the family, we allowed for an arbitrary width of the energy spectrum. If we require that the spectral width does not exceed a given value, the evolution time will be uniformly bounded from below by a positive constant. This is because the energy uncertainty cannot exceed the spectral width.

The Bhatia-Davies inequality \cite{BhDa2000} provides a tighter bound on the energy uncertainty than the spectral width. The Bhatia-Davies inequality states that the variance of any observable $B$ is bounded from above according to 
\begin{equation}\label{Bhatia-Davies ineq}
    \Delta^2 B\leq \langle b_{\mathrm{max}} -  B\rangle\langle B - b_{\mathrm{min}}\rangle,
\end{equation}
with $b_{\mathrm{max}}$ and $b_{\mathrm{min}}$ being the largest and the smallest occupied eigenvalue of $B$. Consequently, the evolution time of an isolated system is bounded by $\tauBD(\delta)$ defined in \eqref{isolatedBD}, and the evolution time of a closed system is bounded by $\tauclBD(\delta)$ defined in \eqref{closedBD}. 

Equality holds in the Bhatia-Davies inequality if and only if the state occupies at most two eigenvalues of $B$. Since the state of an isolated system saturating the Mandelstam-Tamm QSL occupies only two energy levels \cite{Br2003, HoAlSo2022}, the Mandelstam-Tamm and Bhatia-Davies QSLs are always saturated simultaneously for isolated systems.

The Mandelstam-Tamm and Bhatia-Davies QSLs generalize to closed systems as in \eqref{closedMT} and \eqref{closedBD}, respectively, and a natural guess would be that also these QSLs are always saturated simultaneously. However, as we will see, a time-dependent Hamiltonian can evolve a state at a constant speed along a Fubini-Study geodesic in such a way that the state during the entire evolution occupies more than two energy levels. Such an evolution will saturate the Mandelstam–Tamm QSL but not the Bhatia–Davies QSL. This is because the Bhatia-Davies inequality will be strict over the entire evolution time interval, which means that the denominator in \eqref{closedBD} is strictly greater than the denominator in \eqref{closedMT}.

\subsection{A nonsaturation of the Bhatia-Davies quantum speed limit}
Let $H$ be a Hermitian operator with at least three distinct eigenvalues, and let $\rho=\ketbra{u}{u}$ be any state occupying at least three of those. Define $A$ as in \eqref{elegant}, let $H_t=e^{-iAt}He^{iAt}$, and let $\rho_t$ be the state at time $t$ generated from $\rho$ by $H_t$. Since $[H-A,\rho]=0$ and $A\rho+\rho A=A$, the system evolves between two states with fidelity $\delta$ in time $\tauclMT(\delta)$. Furthermore, according to \eqref{occupied}, $\rho_t$ always occupies at least three different energy levels. Therefore,
\begin{equation}
    \Delta^2H_t<\langle\emaxt -  H_t\rangle\langle H_t - \emint\rangle,
\end{equation}
and $\tauclMT(\delta)>\tauclBD(\delta)$. We conclude that the Mandelstam-Tamm QSL is saturated but not the Bhatia-Davies QSL.

\section{Summary}
A common view is that the Margolus-Levitin quantum speed limit extends to an evolution time estimate for closed systems of the form $\mathcal{L}(\delta)/\llangle H_t-\emint\rrangle$, where $\mathcal{L}$ is a positive function that depends only on the fidelity $\delta$ between the initial and final states and $\llangle H_t-\emint\rrangle$ is the time average of the normalized expected energy. We have shown that this is not the case. More precisely, for any fidelity $\delta$ and any positive numbers $E$ and $\mathcal{L}(\delta)$, we have constructed a closed system with a conserved normalized expected energy $E$ that evolves between two states with fidelity $\delta$ in a time strictly less than $\mathcal{L}(\delta)/E$. 

We have also considered a QSL for isolated systems called the Bhatia-Davies QSL. This QSL extends straightforwardly to closed systems. We have shown that the Bhatia-Davies and Mandelstam-Tamm QSLs are always simultaneously saturated for isolated systems but that this need not be the case for closed systems. The state of a closed system that saturates the Mandelstam-Tamm QSL but not the Bhatia-Davies QSL must, at some instant, occupy at least three different energy levels.

\end{document}